\documentclass[10pt,english]{IEEEtran}
\usepackage[T1]{fontenc}
\usepackage{geometry}
\usepackage{amsmath}
\usepackage{xpatch}
\usepackage{amssymb}
\usepackage{esint}
\usepackage{mathtools}
\usepackage{amsthm}
\usepackage{nicefrac}
\usepackage{gensymb}
\usepackage{bigints}
\usepackage{mathtools}
\usepackage{blkarray, bigstrut}
\usepackage{physics}
\usepackage{calligra}
\usepackage{graphicx}
\usepackage{epstopdf}
\usepackage{dsfont}
\usepackage{array,ragged2e}
\usepackage{enumitem}
\usepackage{etoolbox}
\usepackage{babel}
\usepackage[nopar]{lipsum}
\usepackage{psfrag}

\usepackage[ruled,norelsize]{algorithm2e}

\usepackage{balance}
\usepackage{float}
\usepackage{verbatim}
\usepackage[caption = false]{subfig}
\SetKw{KwBy}{by}
\usepackage{algpseudocode}
\usepackage{makecell}
\usepackage{to-be-determined} 

\usepackage[compact]{titlesec}
\usepackage{amsthm}
\newtheorem{definition}{Definition}

\titlespacing{\section}{0.5pt}{*0.5}{*0.5}
\titlespacing{\subsection}{0.5pt}{*0.5}{*0.5}
\titlespacing{\subsubsection}{0.5pt}{*0.5}{*0.5}

\makeatletter
\newcommand{\removelatexerror}{\let\@latex@error\@gobble}
\makeatother

\newtheoremstyle{plain}
  {\topsep}   
  {\topsep}   
  {\itshape}  
  {0pt}       
  {\bfseries} 
  {.}         
  {5pt plus 1pt minus 1pt} 
  {\thmname{#1}\thmnumber{ #2} \textnormal{(\thmnote{#3})}} 

\xpatchcmd{\proof}{\hskip\labelsep}{\hskip5\labelsep}{}{}  
\makeatletter
\xpatchcmd{\proof}{\@addpunct{.}}{\@addpunct{:}}{}{}
\makeatother

\renewcommand\[{\begin{equation}}
\renewcommand\]{\end{equation}} 
\usepackage{listings}
\usepackage{fancyvrb}
\usepackage{framed}

\usepackage{courier}
\definecolor{dkgreen}{rgb}{0,0.3,0}
\definecolor{gray}{rgb}{0.5,0.5,0.5}

\begin{document}

\title{Federated Machine Reasoning for Resource Provisioning in 6G O-RAN}
\author{Swastika Roy$^{(1,2)}$, Hatim Chergui$^{(3)}$, Adlen Ksentini$^{(4)}$  and Christos Verikoukis$^{(5,6)}$\\
{\normalsize{} $^{(1)}$ Iquadrat Informatica S.L.} and {\normalsize{} $^{(2)}$ Technical University of Catalonia (UPC), Barcelona, Spain}\\
{\normalsize{} $^{(3)}$ i2CAT Foundation, Barcelona, Spain}\\
{\normalsize{} $^{(4)}$ Eurecom, Biot, France}\\
{\normalsize{} $^{(5)}$ University of Patras} and   {\normalsize{} $^{(6)}$ ISI/ATHENA, Greece}\\
{\normalsize{}Contact Emails: \texttt{swastika.roy.roy@upc.edu, chergui@ieee.org, adlen.ksentini@eurecom.fr, cveri@ceid.upatras.gr}}}

\maketitle
\thispagestyle{empty}

\begin{abstract}
O-RAN specifications reshape RANs with function disaggregation and open interfaces, driven by RAN Intelligent Controllers. This enables data-driven management through AI/ML but poses trust challenges due to human operators' limited understanding of AI/ML decision-making.
Balancing resource provisioning and avoiding overprovisioning and underprovisioning is critical, especially among the multiple virtualized base station(vBS) instances.
Thus, we propose a novel Federated Machine Reasoning (FLMR) framework, a neurosymbolic method for federated reasoning, learning, and querying. FLMR optimizes CPU demand prediction based on contextual information and vBS configuration using local monitoring data from virtual base stations (vBS) on a shared O-Cloud platform.This optimization is critical, as insufficient computing resources can result in synchronization loss and significantly reduce network throughput.
In the telecom domain, particularly in the virtual Radio Access Network (vRAN) sector, predicting and managing the CPU load of vBSs poses a significant challenge for network operators. Our proposed FLMR framework ensures transparency and human understanding in AI/ML decisions and addresses the evolving demands of the 6G O-RAN landscape, where reliability and performance are paramount.
Furthermore, we performed a comparative analysis using \textit{DeepCog} as the baseline method. The outcomes highlight how our proposed approach outperforms the baseline and strikes a better balance between resource overprovisioning and underprovisioning. Our method notably lowers both provisioning relative to the baseline by a factor of 6.

\end{abstract}


\section{Introduction}
The transition from 5G to 6G networks emphasizes better connectivity and diverse service support. The concept of open radio access networks (RAN) is pivotal in this evolution, with the O-RAN Alliance leading the standardization of Open RAN architecture for interoperability among multi-vendor elements \cite{under_ORAN}.
In O-RAN, AI/ML optimizes CPU resources for efficient network performance across diverse regions with multiple base stations, but data security complexities challenge accurate predictive model development.Transparency and interpretability in predictive models are critical for making informed resource allocation decisions. This preserves the delicate balance between overprovisioning and underprovisioning, ensuring optimal resource utilization, flexibility to adapt to network changes, and cost optimization. 
\cite{vBS-adap} proposes an online learning algorithm for resource allocation balancing throughput and energy consumption, while \cite{baseline} introduces DeepCog for cognitive resource management in 5G, addressing the balance between resource overprovisioning and service request violations.
 In \cite{impact}, the authors evaluate resource allocation in vRAN networks within the O-RAN paradigm, showing that misallocation can harm performance. Given O-RAN's distributed nature, federated reinforcement learning \cite{FL-ORAN} have been proposed for enhancing predictive models while preserving privacy.These studies lack clarity in explaining the interpretability of their predictive models, a common issue with AI/ML adoption in the telco sector.
 Neural networks' limitations in interpretability and systematic generalization emphasize the significance of machine reasoning (MR), a Neural-symbolic AI (NSAI) approach in AI \cite{MR_tech}. In addressing ML limitations, the authors \cite{MR} propose MR as an advanced solution for the growing complexity in telecom networks. 
Thus, we propose integrating  MR with FL to tackle O-RAN challenges. FL enables collaborative learning across base stations, improving AI forecasting with distributed data. This fusion of interpretability from NSAI and privacy-preserving learning of FL builds trust among O-RAN operators and stakeholders.
In this paper, we present the following contributions:
\begin{itemize}
    \item Addressing the challenge within the O-RAN framework by focusing on efficient computing usage of vBS (gNB) instances on shared computing platforms in the O-Cloud.

    \item Developing accurate CPU forecasting models in O-RAN architecture entails addressing data privacy and security and ensuring model transparency for operators to understand high CPU utilization factors.
    Notably, this system-agnostic approach ensures flexibility across O-RAN and non O-RAN environments, enhancing the model's utility for accurate forecasting amidst evolving telecom demands.

    \item Integrating Logical tensor networks (LTN) \cite{LTN}, a NSAI approach, with FL for real-time resource allocation in the Near Real-Time RIC of O-RAN. LTN introduces logical constraints to  provide transparency into decision-making tasks for resource allocation with specific conditions.

     \item Contributing significantly to balance the trade-off between resource overprovisioning and underprovisioning to enhance the efficiency of the network.  

\end{itemize}

\begin{figure}[h]
\centering
    \includegraphics[angle=360,width=70mm, scale=0.30]{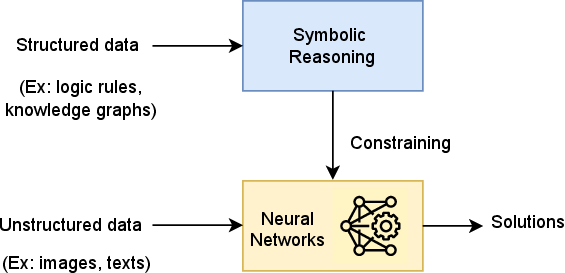}
\caption{Reasoning for learning \cite{survey}}
\label{NN}
\end{figure}

\section{The LTN Paradigm in Machine Reasoning}
Neural-symbolic learning systems combine the strengths of symbolic and neural approaches. Symbolic systems excel with structured data, acquire solution spaces through training and yield higher-level reasoning outputs. Conversely, neural systems show proficiency in learning from unstructured data, producing in lower-level learning outputs during training.  So, by merging both capabilities, it offers a unified technique for problem-solving \cite{survey}. In the \textbf{\textit{reasoning for learning}} category, shown in Fig. \ref{NN}, this approach incorporates symbolic knowledge into training process and leverage the capabilities of neural systems for machine learning tasks.
This enhances interpretability and overall performance in challenging tasks by guiding or constraining the learning process with symbolic knowledge, typically encoded for neural networks. 
 Based on this concept, the paper of \cite{LTN}, introduced LTN which has a notable contribution in this domain. It is neuro-symbolic approach, a MR framework that seamlessly integrates neural and symbolic approaches to learning, reasoning, and querying concrete data and abstract knowledge. 
Using neural computational graphs and first-order fuzzy logic semantics, it presents tensor-based formalism and offers Real Logic, a fully differentiable logical language \cite{LTN}. It introduces logical constructs like predicates and axioms to reasoning and interpretability in machine learning systems. Axioms express general facts by logical assertions, whereas predicates reflect relationships or properties, allowing LTN to hold complex relationships, facts, and rules about entities in a domain. It enables the creation of AI models that anticipate and explain why they predict \cite{LTN}.In telecommunications, transparency is crucial for understanding AI systems' decisions, particularly in architectures like O-RAN. Integrating LTN into the proposed federated deep learning architecture enhances interpretability and aligns with human cognitive processes, vital for dynamic telecom environments.

\begin{figure}[t]
\centering
    \includegraphics[angle=360,width=60mm, scale=0.05]{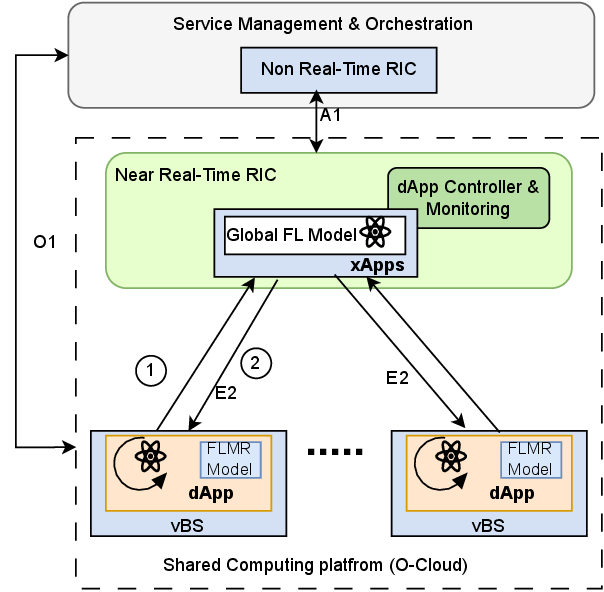}
\caption{Federated model in O-RAN architecture \cite{dApp}}
\label{network}
\vspace{-0.6cm}
\end{figure}

\section{Proposed Architecture and Datasets}
\subsection{Architecture}
As illustrated in Fig. \ref{network}, our O-RAN based system model, drawing inspiration from \cite{ORAN}, involves multiple vBS instances operating on a shared computing platform and establishing connections with User Equipments (UEs). 
The O-RAN system includes two RAN Intelligent Controllers (RICs): a near real-time RIC with vBSs and a non-real-time RIC by SMO. They support programmable functions, leveraging AI, and control activities through xApps and rApps. SMO manages O-Cloud infrastructure via O1/O2 interfaces for vBS setup. Monitoring metrics follow O-RAN standards, with a near-RT RIC retrieving data via E2 and passing it to the non-RT RIC via A1.
Our design expands the traditional O-RAN concept by integrating proactive network resource management into each vBS. Utilizing FL via the Near Real-Time RIC improves mobile communication services' dependability. Independent local FL models in each vBS enhance adaptive resource management, while a Global FL model in the Near Real-Time RIC's xApps ensures system stability and flexibility.
Additionally, the system introduces distributed applications, dApps \cite{dApp}, which complement xApps/rApps by providing intelligence at Central Units (CUs)/Distributed Units (DUs). These dApps support real-time inference at tighter timescales compared to the RICs. In our system model, dApps are situated within the vBS instance. A distinct component, the dApp Controller \& Monitoring, is hosted in the near-RT RIC, controls and monitors dApps executing at the vBSs/gNBs through xApps \cite{dApp}. Noted that, the concept of dApps within the O-RAN framework has been proposed in this work \cite{dApp}, although yet to be standardized by the O-RAN alliance \cite{ORAN}. However,  we have incorporated this entity into our framework due to the unique capabilities of dApps in addressing the specific challenges outlined in our problem statement.

\subsection{Datasets}
The dataset, as summarized in Table I, corresponds to research activities documented in \cite{oran-data}, focusing on the instantiation of varying vBS numbers on a shared computing platform. These vBSs are instantiated in specific CPU core sets with distinct time-sharing allocations. Each vBS is associated with context, encompassing traffic demands and statistics on the used Modulation and Coding Scheme (MCS) for both Uplink (UL) and Downlink (DL). Notably, network parameters like the MCS index significantly impact CPU load, particularly due to coding/decoding workloads. Moreover, this is a  time-series dataset where each row corresponds to a 20-second experiment with specific contextual conditions.It includes per-core CPU utilization, calculated as the average of samples collected every 200 milliseconds, indicating a temporal dimension for analyzing CPU utilization patterns over time.

\begin{table}[h!]
\label{Datasets-tab}
\centering	

\caption{Dataset Features and Output}
{\color{black}\begin{tabular}{lc}
\hline
\hline 

Feature & Description\\
\hline
\texttt{mcs$\_$dl$\_$i}& Downlink MCS index of vBS i \\ 
\texttt{mcs$\_$ul$\_$i} & Uplink MCS index of vBS i\\
\texttt{dl$\_$kbps$\_$i} & Downlink traffic demand in kbps of vBS i\\
\texttt{ul$\_$kbps$\_$i} & Uplink traffic demand in kbps of vBS i\\
\texttt{cpu$\_$set} &  Computing set used by vBS i\\

\hline 
\hline 
& \\
\hline
\hline 
Output & Description\\
\hline
\texttt{cpu$\_$i} & Avg. measured CPU utilization between 0 and 1.\\
\texttt{explode} & Whether the experiment has run correctly or not \\

\hline
\hline
\label{Datasets-tab1}
\end{tabular}
}
\vspace{-0.5cm}
\end{table}

\section{Federated machine reasoning (FLMR) model for transparent CPU usage prediction}
Here, we describe the different stages of the proposed FLMR model as summarized in Fig. \ref{FLMR}.
\subsection{Closed-Loop Description}
\begin{figure}[t]
\centering
    \includegraphics[angle=360, scale=0.70]{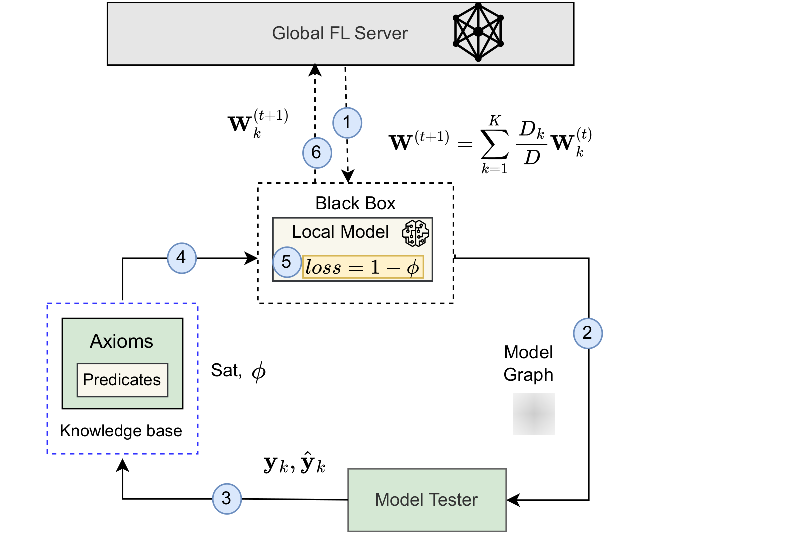}
    \caption{Neuro symbolic FLMR block}
\label{FLMR}
\vspace{-0.5cm}
\end{figure}

We propose a federated deep learning architecture where the local learning is performed iteratively with run-time rule-based neuro symbolic reasoning scheme in a closed loop way as shown in Fig. \ref{FLMR}. Our model integrates LTN \cite{LTN}, capable of both neural network-style learning and classical AI-style reasoning, into a deep neural network FL model.
For each local epoch, the Learner module feeds the posterior symbolic model graph to the Tester block which yields the test features and the corresponding predictions $\hat{y}_{k}^{(i)}$ to the \emph{Knowledge base Mapper} at step 3. In the proposed ORAN based system model, this function should be deployed in the xApps. xApps are responsible for implementing specific functions or services and are part of the broader move towards a more open, interpretable and flexible RAN environment. 
During training, LTNs calculate the satisfaction level of logical constraints from the Knowledge base information at step 4. The \textbf{satisfaction level}, defined as $\phi$, is converted into a loss during training using the formulation "$loss = 1 - \phi$." In the training process, the optimization algorithm, such as the Adadelta optimizer, works to minimize this loss at step 5. It adjusts the model's parameters, including weights and biases, to simultaneously maximize satisfaction of logical constraints and minimize the loss. The objective is to find optimal parameter values that strike a balance between accurate predictions and adherence to the logical rules encoded in the "\textbf{\textit{eq}}" predicate which we explain in the next subsection.

Indeed, for each local CL $(k)$, the predicted ML model $\hat{y}_{k}^{(i)},\,(i=1,\ldots,D_{k})$, should minimize the main loss function with respect to the ground truth $y_{k}^{(i)}$, while jointly respecting some long-term logical constraints defined over its $D_{k}$ samples.

As shown in steps 1 and 6 of Fig. \ref{FLMR}, the optimized local weights at round $t$, $\mathbf{W}_{k}^{(t)}$, are sent to the server which generates a global FL model as,
\begin{equation}
\label{GlobalFL}
    \mathbf{W}^{(t+1)}=\sum_{k=1}^K \frac{D_{k}}{D}\mathbf{W}_{k}^{(t)},
\end{equation}
where $D=\sum_{k=1}^K D_{k}$ is the total data samples of all datasets. The server then broadcasts the global model to all the $K$ CLs that use it to start the next round of iterative local optimization. 

\subsection{Knowlegde base Mapper}
However, in LTN, the knowledge base is a central component that stores and represents information in a formalized manner. The knowledge base in LTN typically consists of predicates, axioms, and other logical constructs that capture relationships, facts, and rules about the entities in a given domain.
It forms the basis for logical reasoning and decision-making within the framework, allowing for the representation and manipulation of complex knowledge structures. 
\subsubsection{Predicates}
\emph{Predicates} in the knowledge base represent relationships or properties between entities. We define our predicates by inspiring the paper of \cite{LTN} which introduce a  "smooth" version of the equality symbol "$=$" using a predicate called "\textbf{\textit{eq}}." 
The "\textbf{\textit{eq}}" predicate is defined as: 
\begin{equation}
\textbf{\textit{eq}}(f(x_i), y_i) = \frac{1}{1 + \alpha \sqrt{\sum_{j=1}^{N} (f(x_i)_j - (y_i)_j)^2}}
\end{equation}
Here, $\textbf{\textit{eq}}(f(x_i), y_i)$ represents the output of the "eq" predicate for predicted value $f(x_i)$ and observed value $y_i$. $\alpha$ is a positive constant that controls the smoothness of the matching.
 This formulation ensures output within the [0, 1] range, with higher values indicating a closer matches.
In our O-Cloud scenario, the "\textbf{\textit{eq}}" predicate provides crucial domain-specific knowledge by allowing for flexible matching between predicted ($f(x_i)$) and observed ($y_i$) CPU usage in a continuous and smooth manner. This flexibility is essential for optimizing vRAN network performance in O-RAN deployment, as traditional strict equality constraints can be overly rigid, making optimization challenging. 

\subsubsection{Axioms}
\emph{Axioms} are logical statements or rules that express general truths within the knowledge base, guiding the process of making inferences based on the predicates' relationships. The logical relationships specified by predicates within axioms enable the system to derive new information.
During training, LTNs calculate the satisfaction level of logical constraints from the Knowledge base information, shown at step 4. This satisfaction level quantifies how well the model adheres to the specified logical rules, such as the "\textbf{\textit{eq}}" predicate, during predictions. 

A high satisfaction level (close to 1) corresponds to a low loss (close to 0), adherence to logical constraints. Conversely, a low satisfaction level (close to 0) results in a high loss (close to 1).

\begin{definition}[\textbf{Satisfaction level}]
   A continuous parameter called the satisfaction level ($\phi$) of logical constraints in LTNs indicates how closely the model complies with predefined logical rules, like the "\textbf{\textit{eq}}" predicate, throughout training. Based on the data from the Knowledge base at a certain phase, this measure is calculated. Typically, $\phi$ ranges from 0 to 1, with higher values denoting a more crucial adherence to the logical requirements.
\end{definition}

\subsection{Integration of LTN with FL for CPU Usage prediction}
In order to solve a local regression problem for predicting CPU usage in iterations specified by the FL rounds $t\,(t=0,\ldots,T-1)$ i.e.,

\label{OPT1}
\begin{equation}
    \min_{\mathbf{W}_{k}^{(t)}}\, \frac{1}{D_{k}}\sum_{i=1}^{D_{k}}\ell\left(y_{k}^{(i)}, \hat{y}_{k}^{(i)}\left(\mathbf{W}_{k}^{(t)},\mathbf{x}_{k}\right)\right),
\end{equation}

Indeed, for each local CL $(k)$, the predicted ML model $\hat{y}_{k}^{(i)},\,(i=1,\ldots,D_{k})$, should minimize the main loss function with respect to the ground truth $y_{k}^{(i)}$, while jointly respecting some long-term logical constraints defined over its $D_{k}$ samples.

The loss is computed as $1-\phi$, where $\phi$ is the satisfaction level of the axioms. Therefore, the loss function, $\ell({y}_{k}^{(i)},\hat{y}_{k}^{(i)})$ can be written as:

\begin{equation}
    \ell({y}_{k}^{(i)},\hat{y}_{k}^{(i)}) = 1 - \phi({y}_{k}^{(i)},\hat{y}_{k}^{(i)})
\end{equation}
The goal of training is to minimize this loss, which essentially maximizes the satisfaction level of the logical axioms.
The axioms are given by:
\begin{equation}
    \textbf{A}({y}_{k}^{(i)},\hat{y}_{k}^{(i)}) = \textbf{\textit{Forall}}(\textbf{\textit{ltn.diag}}({y}_{k}^{(i)},\hat{y}_{k}^{(i)}), \textbf{\textit{eq}}({y}_{k}^{(i)},\hat{y}_{k}^{(i)}))
\end{equation}
This equation represents a logical statement that involves quantifiers (universal quantification with \textbf{\textit{Forall}}), a diagonal operator (\textbf{\textit{diag}}), an equality predicate (\textbf{\textit{eq}}), and the regressor function $f(x)$. The overall expression asserts that for all pairs of variables ${y}_{k}^{(i)}$ and $\hat{y}_{k}^{(i)}$, the equality predicate should hold between $f(x)$ and ${y}_{k}^{(i)}$. The \textbf{\textit{Forall}} quantifier signifies that this condition should be true for all possible combinations of
${y}_{k}^{(i)}$ and $\hat{y}_{k}^{(i)}$.
Additionally, a regression task requires a notion of equality. We, therefore, define the predicate \textbf{\textit{eq}} as a smooth version of the symbol "$=$" to turn the constraint $\hat{y}_{k}^{(i)} = {y}_{k}^{(i)}$ into a smooth optimization problem. The overall steps are mentioned in the Algorithm 1.

\algblock{ParFor}{EndParFor}
\algnewcommand\algorithmicparfor{\textbf{parallel for}}
\algnewcommand\algorithmicpardo{\textbf{do}}
\algnewcommand\algorithmicendparfor{\textbf{end parallel for}}
\algrenewtext{ParFor}[1]{\algorithmicparfor\ #1\ \algorithmicpardo}
\algrenewtext{EndParFor}{\algorithmicendparfor}

\begin{algorithm}[t]
\caption{Federated Machine Reasoning \& Learning}
\footnotesize

\SetAlgoLined
\KwIn{$K$, $\eta_{\lambda}$, $T$, $L$, $\alpha$. \texttt{\# See Table II}}
Server initializes $\mathbf{W}^{(0)}$ and broadcasts it to the 
$\mathrm{CL}$s\\

\texttt{\# Federated Learning}\\
\For{$t=0,\ldots,T-1$}{
\textbf{parallel for} $k=1,\ldots,K$ \textbf{do} \\
\texttt{\# Model graph from the local Model} \\
Receive the graph $\mathcal{M}_{k}$ from the local model. \\
\texttt{\# Test local model and send results to Knowledge Base Mapper.}\\
\texttt{\# Use "\textit{eq}" predicate to measure the satisfaction level}\\

$\textbf{\textit{eq}}(f(x_i), y_i) = \frac{1}{1 + \alpha \sqrt{\sum_{j=1}^{N} (f(x_i)_j - (y_i)_j)^2}}$\\

\texttt{\# Calculate the loss function} \\

{$\ell({y}_{k}^{(i)},\hat{y}_{k}^{(i)}) = 1 - \phi({y}_{k}^{(i)}\hat{y}_{k}^{(i)})$} \\

\texttt{\# Adadelta optimizes to minimize loss, maximizing satisfaction of logical axioms.}\\

 $\textbf{A}({y}_{k}^{(i)},\hat{y}_{k}^{(i)}) = \textbf{\textit{Forall}}(\textbf{\textit{ltn.diag}}({y}_{k}^{(i)},\hat{y}_{k}^{(i)}), \textbf{\textit{eq}}({y}_{k}^{(i)},\hat{y}_{k}^{(i)}))$

  Each local $\mathrm{vBS}$ $k$ sends $\mathbf{W}_{k}^{(t)}$ to the aggregation server. \\
 \textbf{end parallel for}\\
 \texttt{\# FL Server Aggregation}\\
 \Return{$ \mathbf{W}^{(t+1)}=\sum_{k\in\{k_1,\ldots,k_m\}}\frac{D_{k}}{D}\mathbf{W}_{k}^{(t)}$}\\
 Broadcasts $\mathbf{W}^{(t+1)}$ to all $K$ $\mathrm{vBS}$s.}
\end{algorithm}

\begin{table}[htb]
\label{Table1}
\centering	
\newcolumntype{M}[1]{>{\centering\arraybackslash}m{#1}}
\caption{Settings}
\begin{tabular}{ccc}
\hline 
\hline
Parameter & Description & Value\\
\hline
$MLP$ & Multilayer Perception (ANN) & 2 hidden layers\\
$K$ & \# vBSs & $50$\\ 
$D_{k,n}$ & Local dataset batch size & $500$ samples\\ 
$T$ & \# Max FL rounds & $50$\\  
$\eta_{\lambda}$ & Learning rate (AdaDelta) & $0.85$ \\ 
$\alpha$ & Positive constant (\textbf{"eq"} predicate) & $0.5$ \\ 
\hline
\hline
\label{FLsettings}
\end{tabular}
\vspace{-0.8cm}
\end{table}

\section{Results}
This section analyses the proposed closed loop FLMR framework in detail. To build the logical reasoning based predictive model, we include logical reasoning and learning into  local FL model training.

\subsection{Parameter Settings and Baseline}
\begin{itemize}
    \item Settings: For simulations, we utilized Python in an Ubuntu 20.04 environment on a laptop. The datasets, sourced from vBSs, are detailed in Table I. The hyperparameters of our FLMR model are listed in Table II, with AdaDelta as the optimizer.
    \item Baseline:
    As a baseline, we employ the \emph{DeepCog} framework \cite{baseline}, which introduces a novel data analytics tool for the cognitive management of resources in 5G systems. It introduces a customized loss function designed for capacity forecasting, enabling operators to balance overprovisioning and demand violations effectively. This function considers the costs associated with underprovisioning and overprovisioning, ensuring optimal resource allocation.

\end{itemize}

In the following section, we conduct a comparative analysis between our proposed FLMR model and the established deepCog model, evaluating aspects such as convergence and satisfiability. Additionally, we explore the trade-off between over and under resource provisioning in our analysis.

\begin{figure}[htb]
    \centering
    \includegraphics[angle=360,width=80mm, scale=0.65]{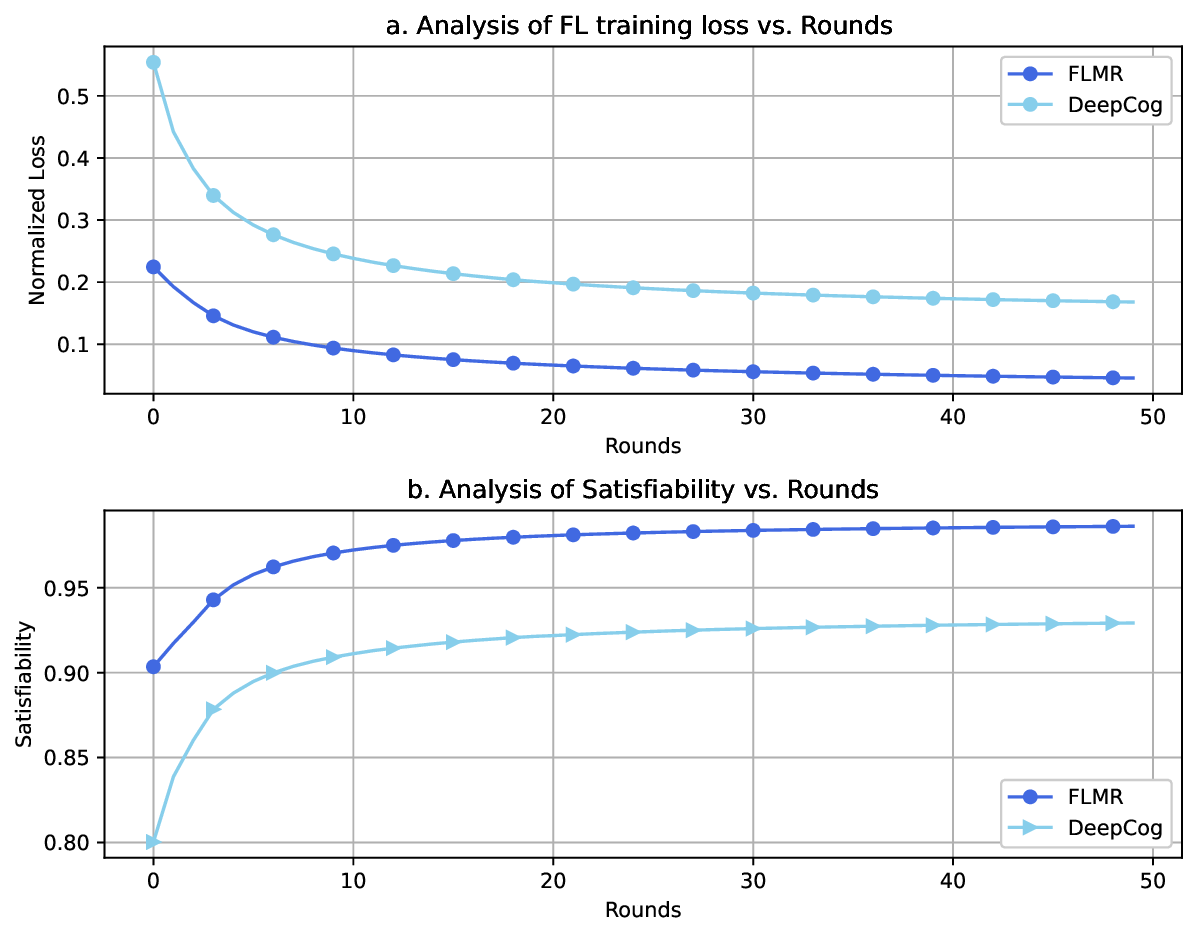}
\caption{Comparative performance analysis between the proposed FLMR and baseline DeepCog FL}
\label{Performance}
\end{figure}

\begin{figure}[htb]
    \centering
    \includegraphics[angle=360,width=80mm, scale=0.65]{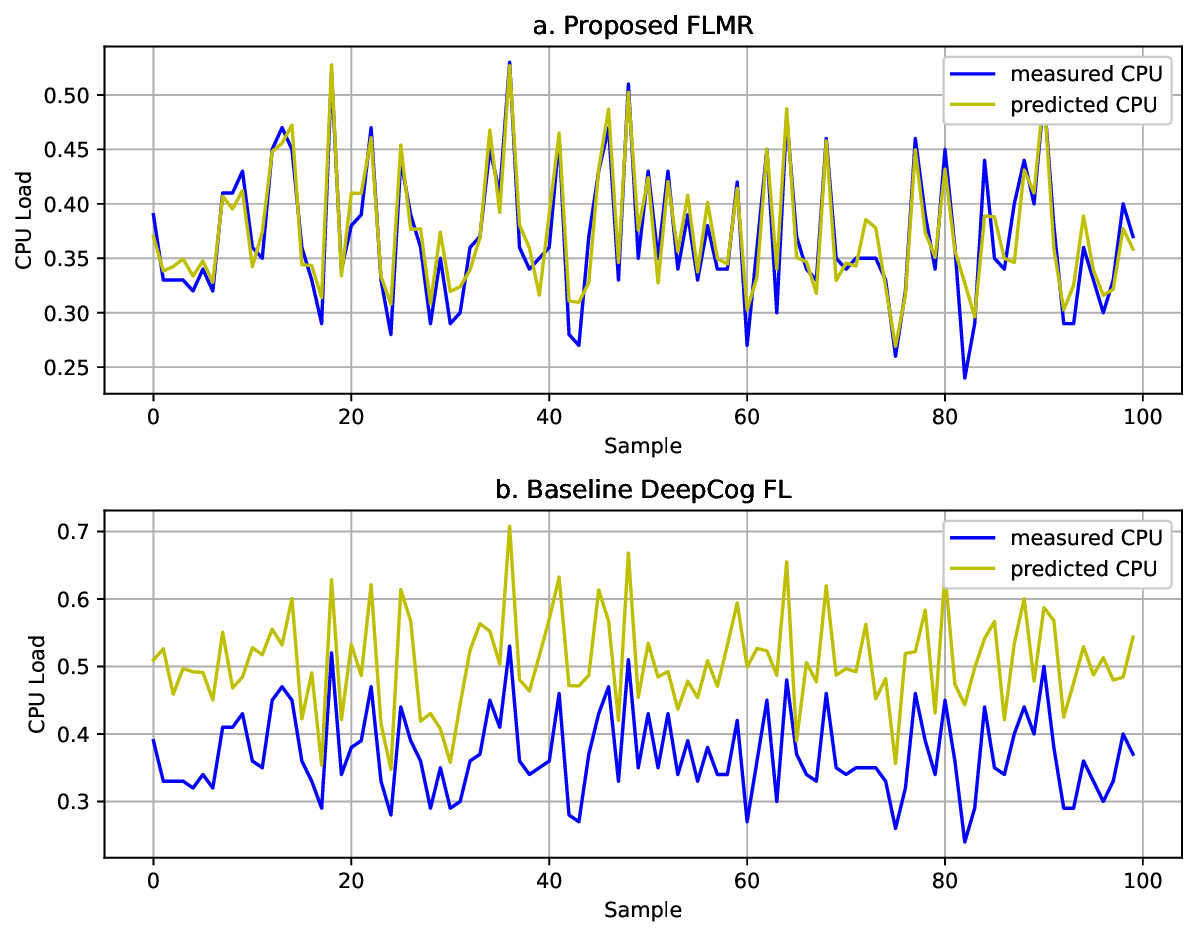}
\caption{Comparative Analysis of CPU Usage Load Prediction: Original vs. Predicted CPU Load}
\label{prediction}
\vspace{-0.5cm}
\end{figure}

\begin{figure*}[htb]
    \centering
    \subfloat[ \centering FLMR]{
    \label{MR_pro}
          \includegraphics[width=0.43\textwidth]{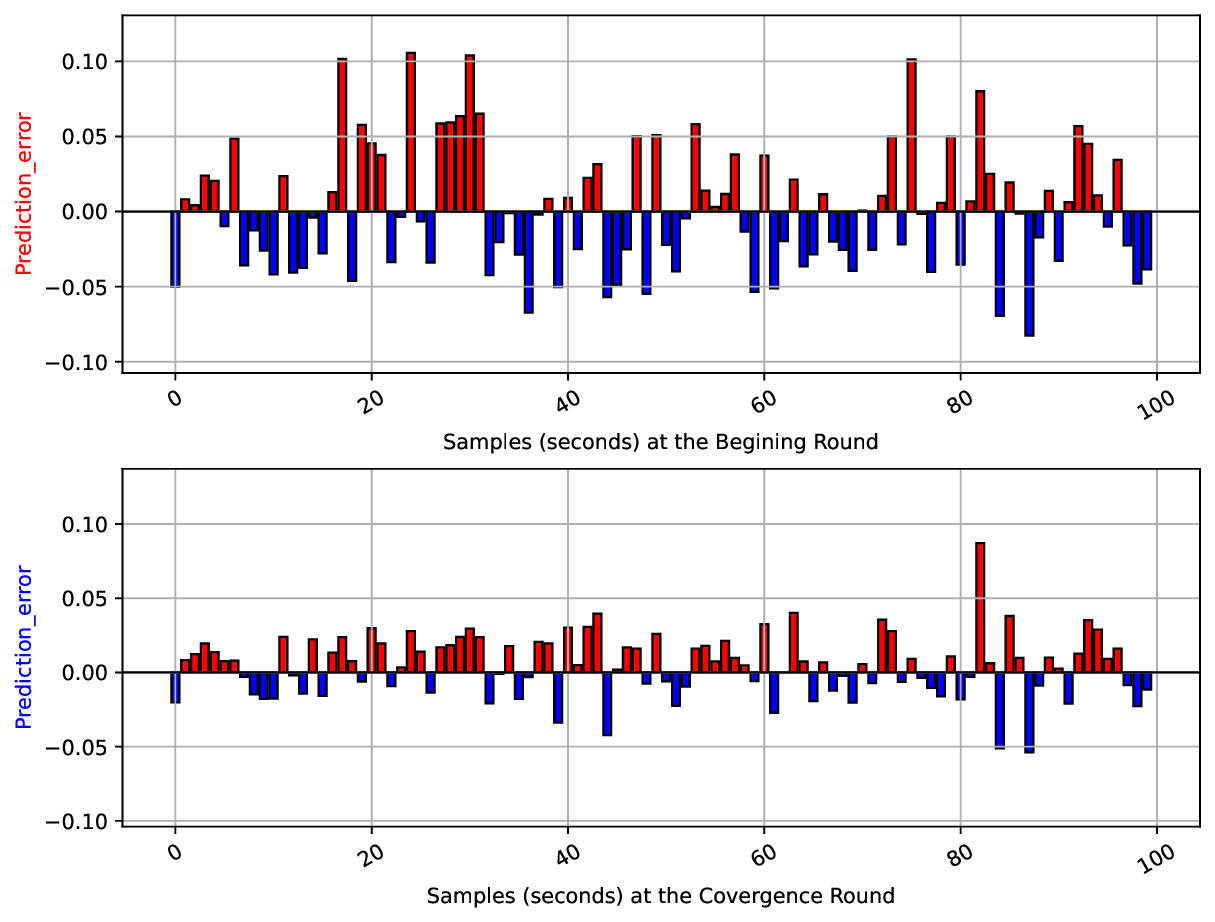}\hspace{1.2cm}}
    \qquad 
    \subfloat[ \centering DeepCog FL]{
    \label{Deep_pro}
          \includegraphics[width=0.43\textwidth]{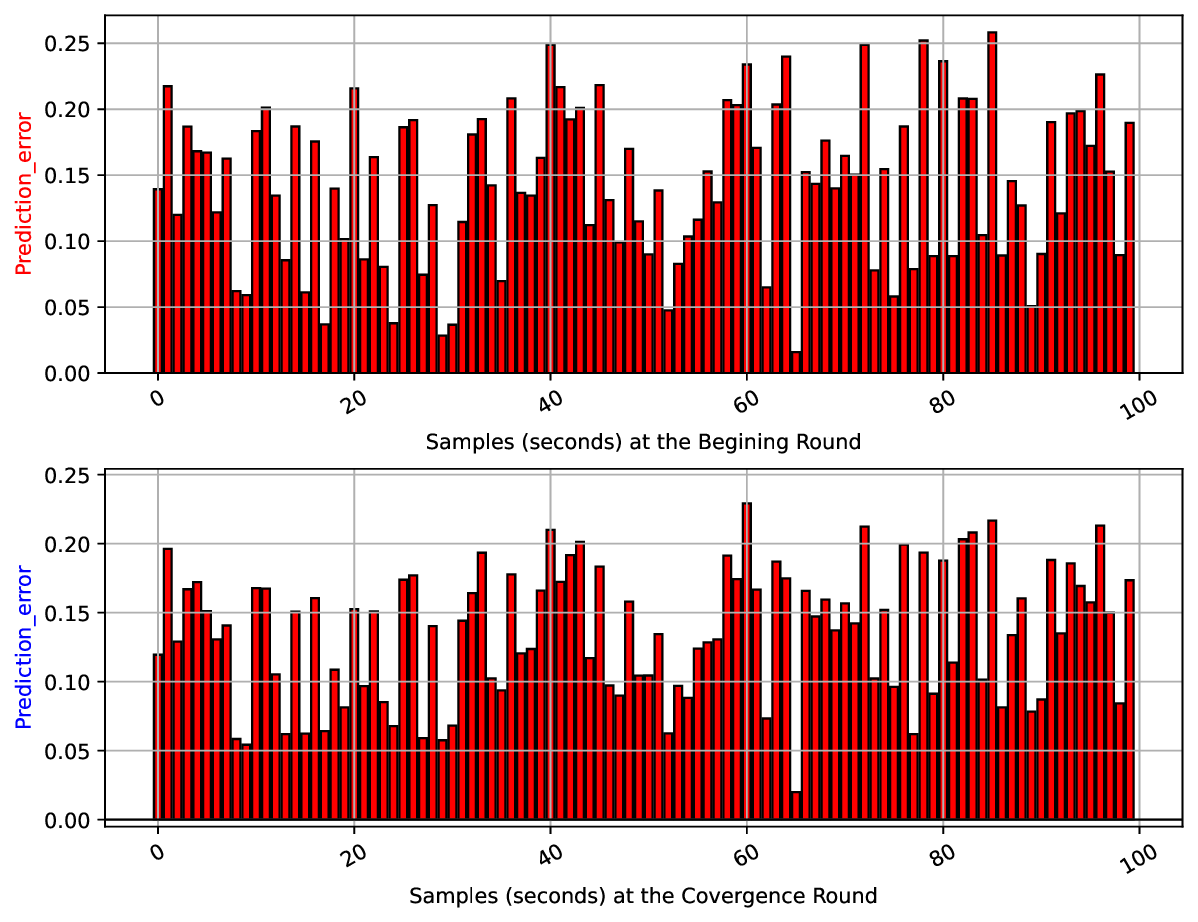}}
\caption{Trade-off between the  resource overprovisioning (Red level) and underprovisioning (Blue level)}
\label{trade-off}
\vspace{-0.5cm}
\end{figure*}

\subsection{Results Analysis}
\subsubsection{Performance Analysis:}
Fig. \ref{Performance}a shows that FLMR achieves faster convergence than the DeepCog FL baseline model. 

The satisfaction levels over FL rounds of both models are shown in Fig. \ref{Performance}b. The satisfaction level serves as a metric to gauge the model's effectiveness and user contentment during the FL training process.
Remarkably, the FLMR model attains a highly desirable satisfaction level, hovering around 98$\%$. This outcome indicates better performance and user acceptance. By comparison, the baseline DeepCog FL model exhibits a comparatively lower satisfaction level.
FLMR's superiority can be attributed to its integration of logical reasoning through LTNs, which effectively handle logical constraints while optimizing the trade off between resource allocation and SLA violations. In contrast, DeepCog FL primarily focuses on capacity forecasting and may not explicitly consider logical constraints, potentially leading to less accurate predictions.

\subsubsection{Validation of Predictive Models: CPU Usage Evaluation}
Fig. \ref{prediction} compares the proposed model's accuracy in predicting CPU load usage to the baseline model. The proposed model, as shown in Fig.  \ref{prediction}a, closely aligns with actual CPU load, showcasing its effectiveness. In contrast, the baseline model, as exhibits in Fig. \ref{prediction}b, struggles to make accurate predictions, highlighting its limitations. The superior performance of the FLMR model is attributed to its unique design choices, including FL for collaborative learning, NSAI for interpretability, and adaptive resource management for efficient responses to CPU demand fluctuations.

\subsubsection{Overprovisioning and Underprovisioning trade-off analysis}
In Fig. \ref{trade-off} analyzes the trade-off between resource overprovisioning and underprovisioning in the proposed FLMR model compared to the baseline. The figure displays prediction errors for 100 samples at the start and end of the FL round.
The prediction error is the difference between the predicted output $\hat{y}_{k}^{(i)}$ and measured  ${y}_{k}^{(i)}$. We can define it by,
\begin{equation}
   P_{err}(x, y) = \hat{y}_{k}^{(i)}- {y}_{k}^{(i)}   
\end{equation}
The red part indicates overprovisioning, where predicted resources exceed actual needs, while the blue part represents underprovisioning, risking SLA violations due to fewer allocated resources. 

The presented model in Fig. \ref{MR_pro} effectively balances the trade-off, minimizing both over-provisioning and under-provisioning during FL convergence. This surpasses the \textit{DeepCog} baseline model shown in Fig. \ref{Deep_pro} by a factor of 6.

\section{Conclusion}
In conclusion, we present a unique neurosymbolic-based FLMR framework that overcomes the difficulty of comprehending AI/ML decision processes in vBS instances while maintaining a careful balance between overprovisioning and underprovisioning resources. By optimizing CPU demand for virtual base stations in the dynamic 6G O-RAN landscape, FLMR stands out as a transparent and optimized solution. The comparative analysis underscores FLMR's superiority in achieving a more effective equilibrium between resource overprovisioning and underprovisioning. FLMR presents itself as a state-of-the-art option for transparent and effective AI/ML decision-making in the dynamic O-RAN ecosystem as the telecom sector develops.

\section{Acknowledgment}


\vspace{1mm}
\bibliographystyle{IEEEtran}
\bibliography{myBibliographyFile}

\balance

\end{document}